\documentclass[amssymb,prd,aps,amsmath,superscriptaddress,nofootinbib,twocolumn,preprintnumbers]{revtex4}

\usepackage{color}
\usepackage{pslatex}
\usepackage{graphicx}
\usepackage{psfrag}
\usepackage{graphicx}
\usepackage{multirow}
\usepackage[utf8]{inputenc}
\usepackage{url}

\makeatletter
\def\url@leostyle{%
  \@ifundefined{selectfont}{\def\UrlFont{\sf}}{\def\UrlFont{\small\ttfamily}}} 
\makeatother
\begin{document}

\title{Observing Higgs boson production through its decay into
  $\gamma$--rays: \\ 
A messenger for Dark Matter candidates}

\preprint{Bonn-TH-2012-030}
\preprint{IPPP/12/85}
\preprint{DCPT/12/72}
\preprint{CFTP/12-017}

\author{Nicol\'as Bernal}
\affiliation{Bethe Center for Theoretical Physics and
  Physikalisches Institut, Universit\"at Bonn, Nu\ss allee 12, D-53115
  Bonn, Germany}

\author{C\'eline B\oe hm}
\affiliation{Institute for Particle Physics Phenomenology, University
  of Durham, Durham, DH1 3LE, UK} 
\affiliation{LAPTH, U. de Savoie, CNRS,  BP 110,
  74941 Annecy-Le-Vieux, France.}

\author{Sergio Palomares-Ruiz}
\affiliation{Centro de F\'isica Te\'orica de Part\'iculas (CFTP),
Instituto Superior T\'ecnico, Universidade T\'ecnica de Lisboa, Av.
Rovisco Pais 1, 1049-001 Lisboa, Portugal}
\affiliation{Instituto de F\'{\i}sica Corpuscular (IFIC),
  CSIC-Universitat de Val\`encia, Apartado de Correos 22085, E-46071
  Valencia, Spain} 

\author{Joseph Silk}
\affiliation{UMR7095 - Institut d'Astrophysique de Paris - 98 bis
  boulevard Arago - 75014 Paris} 

\author{Takashi Toma}
\affiliation{Institute for Particle Physics Phenomenology, University
  of Durham, Durham, DH1 3LE, UK}

\begin{abstract}
In this Letter, we study the $\gamma$--ray signatures subsequent to
the production of a Higgs boson in space by dark matter annihilations.
We investigate the cases where the Higgs boson is produced at rest or
slightly boosted and show that such configurations can produce
characteristic bumps in the $\gamma$--ray data.  These results are
relevant in the case of the Standard Model-like Higgs boson provided
that the dark matter mass is about 63~GeV, 109~GeV or 126~GeV, but can
be generalised to any other Higgs boson masses.  Here, we point out
that it may be worth looking for a 63~GeV line since it could be the
signature of the decay of a Standard Model-like Higgs boson produced
in space, as in the case of a di-Higgs final state if $m_{\chi} \simeq
126$~GeV.  We show that one can set generic constraints on the Higgs
boson production rates using its decay properties.  In particular,
using the {\sl Fermi}--LAT data from the galactic center, we find that
the dark matter annihilation cross section into $\gamma \, +$ a Standard
Model-like Higgs boson produced at rest or near rest cannot exceed
$\langle \sigma \, v \rangle \sim \rm{a \ few} \, 10^{-25}
\rm{cm^3/s}$ or $\langle \sigma \, v \rangle \sim \rm{a \ few} \,
10^{-27} \rm{cm^3/s}$ respectively, providing us with information on
the Higgs coupling to the dark matter particle.  We conclude that
Higgs bosons can indeed be used as messengers to explore the dark
matter mass range.
\end{abstract}

\date{today}

\maketitle

\section{Introduction}
On-going searches at the LHC have been rewarded by one of the greatest
particle physics discoveries that could possibly be  made in such a
machine, namely the finding of a seemingly new fundamental scalar or
pseudo-scalar particle~\cite{Chatrchyan:2012tw, Chatrchyan:2012tx,
  ATLAS:2012ad, ATLAS:2012ae}.  At present measurements of the
couplings of this new boson to Standard Model (SM) particles along
with the absence of charged particles tend to suggest that this is a
SM Higgs boson.  However this remains to be proven.  

While such a discovery certainly validates our understanding of the
origin of particle masses, it also constrains the types of theories
that could be proposed to go beyond the Standard Model (BSM).  For
instance, some of the simplest Supersymmetric (SUSY) models which have
been proposed in the literature tend to predict a mass for the Higgs
boson that is smaller than the measured value $m_H \simeq
125-126$~GeV~\cite{Arbey:2012dq} and are therefore likely to be ruled
out.  Moreover, the good agreement between the measured branching
ratios and those expected in the SM (apart perhaps for the two-photon
channel) enables one to set a stringent constraint on the Higgs
invisible decay width and to constrain theories in which the 
Higgs is strongly coupled to the dark matter (DM) candidate
  ($\chi$)~\cite{Englert:2011us}.

Nevertheless, the information collected so far at the LHC is not
sufficient to exclude the possibility that this new boson has a BSM
origin.  In fact, some non-minimalistic SUSY extensions were shown to
predict a `light' Higgs boson with essentially indistinguishable
characteristics from those expected within the SM (the remainder of
the spectrum in this model being typically beyond the scale accessible
at LHC)~\cite{Vasquez:2012hn}.  Hence, at present the origin of this
new boson remains an open question and one needs more clues to
determine whether this Higgs boson candidate has a SM origin or not.
Examining its `dark' coupling using other tools than the LHC could be
one way to proceed.

In this Letter, we  propose to exploit this discovery together
with recent astrophysical data to constrain the Higgs boson production
cross section in some specific annihilating DM scenarios. We shall
focus on the SM-like boson with a mass of 126~GeV, but our analysis can
be extended to any Higgs boson candidate.  Now that a SM-like Higgs
(or a new) boson has been discovered and its main characteristics are
well determined, one can make use of its decay properties (and in
particular the photon spectrum subsequent to the Higgs boson decay) to 
determine whether it has been produced by DM in our galactic halo, for
instance.  Observing the decay of a Higgs boson  produced at rest (or
slightly boosted) in space would indeed be suggestive of new physics
and provide a new window on long-lived neutral particles. The scheme
that we have in mind is the production of one or two Higgs bosons by
DM annihilations, although an analogous exercise can be done for
decaying DM, with similar qualitative arguments for DM masses a
factor of 2 higher.  Once a Higgs boson is produced, it is expected to
decay immediately, thereby  generating $\gamma$--rays.  If the
associated flux is large enough, this could lead to anomalous features
in the $\gamma$--ray spectrum (in particular, an excess of photons at
some specific energies with respect to the background expectations)
which can be searched for.  Note that in what follows we will only
focus on the  $\gamma$--ray emission from the galactic centre, but our
analysis could be extended to other regions of the Milky Way as well
as the emission arising from DM annihilations in dwarf galaxies.

The $\gamma$--ray signature associated with a SM-like Higgs boson
decay in our galaxy is expected to be a smooth continuum spectrum due
to the Higgs decay into SM particles~\cite{Jackson:2009kg}.  However,
here we show that if the Higgs boson is produced at rest, its decay
into two gamma ($H \rightarrow \gamma \, \gamma$) could lead to a
potentially detectable monochromatic line at $E_{\gamma}\sim63$~GeV in
addition to the continuum, even though the associated branching ratio
is very suppressed with respect to other channels. 
 
The corresponding signal in an experiment such as {\sl Fermi}--LAT
should be a bump around $E_{\gamma} =m_H/2$ (that is
$E_{\gamma}\sim63$~GeV for a SM-like Higgs boson) and possibly a broad
$\gamma$--ray excess at lower energies, depending on the ratio between
the line and the continuum.  Here we show that it is worth looking for
such a line in $\gamma$--ray data, as it could be a mean to probe
specific annihilating DM scenarios.  In particular, in the case of a
SM-like Higgs boson, one could probe DM masses of about $m_\chi \simeq
63$~GeV (for $\chi \, \chi \rightarrow H \, \gamma$), $m_\chi \simeq
109$~GeV (for $\chi \, \chi \rightarrow H \, Z$) or $m_\chi \simeq
126$~GeV (for $\chi \, \chi \rightarrow H \, H$)~\footnote{For the DM
  masses under consideration, the DM-induced synchrotron signal lies
  at radio frequencies.  Present radio data from the galactic center
  could also give constraints on DM properties~\cite{Laha:2012fg}.}. 

In Section~\ref{sec:rest} we discuss the production of the SM-like
Higgs boson at rest in DM annihilations.  After reviewing the possible
DM annihilation processes which can create one (or two) Higgs boson(s)
in the final state, we study the detectability of the signature of a
Higgs boson decay with the Large Area Telescope ({\sl Fermi}--LAT) on
board the {\sl Fermi} mission and discuss the implications for DM
scenarios.  We also comment on the slightly boosted Higgs boson in
Section~\ref{sec:boost} and conclude in Section~\ref{sec:conclusions}.

\section{Higgs boson produced at rest by DM
  annihilations \label{sec:rest}}

\begin{figure}[t]
\centering
\includegraphics[width=0.24\linewidth]{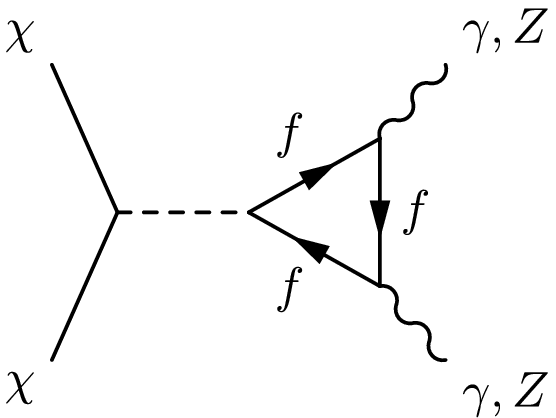}
\includegraphics[width=0.24\linewidth]{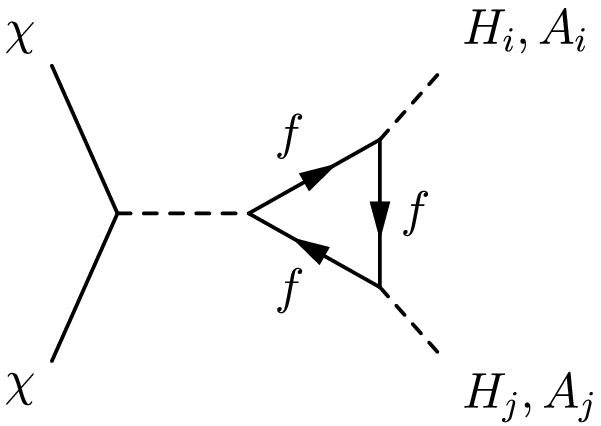}
\includegraphics[width=0.24\linewidth]{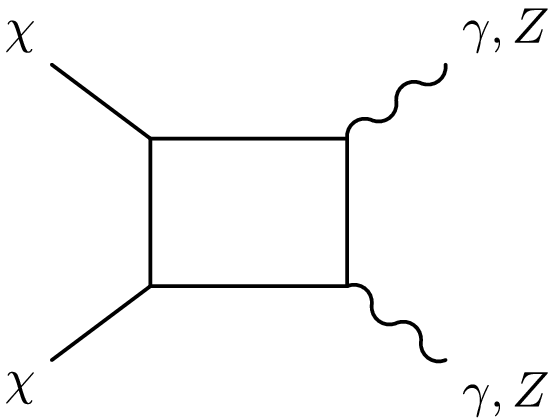}
\includegraphics[width=0.24\linewidth]{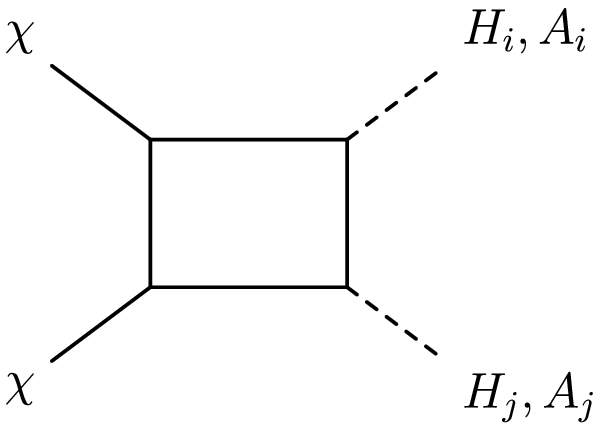}
\includegraphics[width=0.24\linewidth]{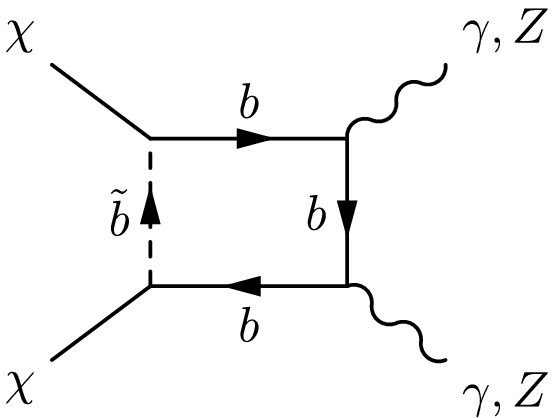}
\includegraphics[width=0.24\linewidth]{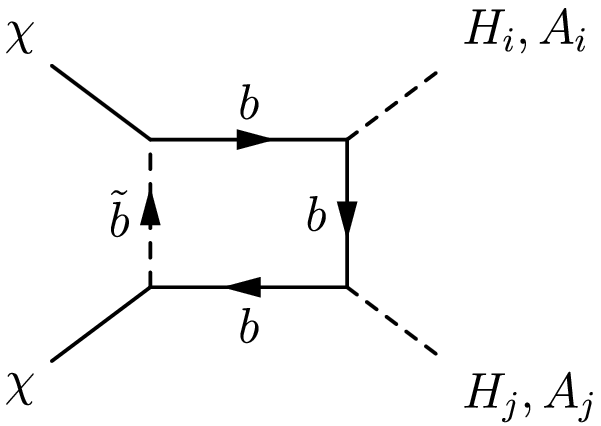}
\includegraphics[width=0.24\linewidth]{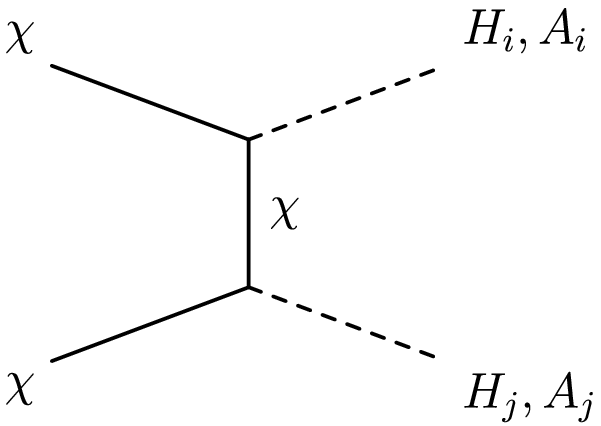}
\includegraphics[width=0.24\linewidth]{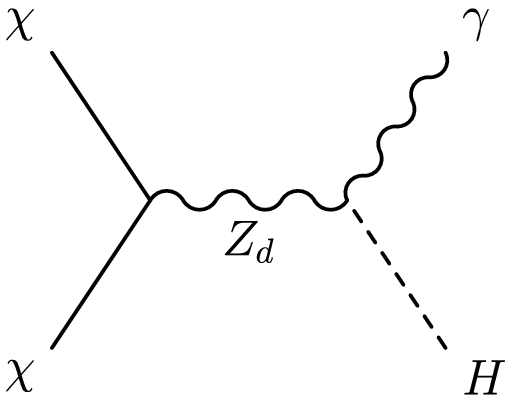}
\caption{Examples of Feynman diagrams associated with relevant process
  discussed in this paper.}
\label{diagram}
\end{figure}

In order to produce a Higgs boson in space and at rest, the DM mass
and spin must have specific values.  Quantitative statements depend on
how many Higgs bosons are produced in the final state.  In the case of
DM annihilations into two SM-like Higgs bosons, the DM mass must be
about $m_\chi \simeq m_H \simeq$~126~GeV (regardless of its spin).  If
on the contrary, DM annihilations produce only one SM-like Higgs boson
plus a photon in the final state, the DM mass must be about $m_\chi
\simeq m_H/2 \simeq 63$~GeV (assuming that it has a spin-1/2 or
spin-1) while it should be about 109~GeV if it produces a Higgs boson
plus a $Z$ boson in the final state (assuming a spin-0,1/2 or spin-1).
In what follows, we will focus on these three specific cases, as they
lead to the production of SM-like Higgs bosons at rest but, of course,
an analogous analysis can be done for heavier (presumably BSM) Higgs
bosons.  We now point out some general Higgs boson production
mechanisms which could prevail for DM candidates with a mass $m_\chi
\simeq$~63~GeV, 109~GeV and 126~GeV.  Examples of relevant Feynman
diagrams are given in Fig.~\ref{diagram}.

\subsection{Production mechanisms for $m_\chi \simeq 126$~GeV}

DM candidates with a mass slightly greater than 126~GeV can produce
two Higgs bosons at rest or near rest in the final state either
through box diagrams or, if DM is directly coupled to the Higgs,
through tree-level process (see Fig.~\ref{diagram}). 

In a SUSY framework for example, two Higgs bosons can be produced via
box diagrams involving, e.g., charginos and $W$ boson or quarks and
squarks from the third generation~\cite{Bergstrom:1997fh}.  Disregarding
for the moment the possible velocity-squared dependence which arises due
to the Majorana nature of the neutralino, these diagrams are expected
to be relatively suppressed with respect to other annihilation
channels which occur at tree-level (such as for example neutralino
annihilations into $b \, \bar{b}$ or $W^+ \, W^-$ via a t--channel
sbottom or chargino exchange respectively).  However they could still
be sizable if the Higgs boson has large couplings to the
particles in the box or if there is a large mass degeneracy between
the neutralino and the chargino ($\chi^\pm$) for example (if we consider
the $\chi^\pm-W^\mp$ box diagram~\cite{Bergstrom:1997fh, Bern:1997ng, 
  Boudjema:2005hb, Belanger:2012ta}).  Alternatively, the DM could
also pair annihilate into two Higgs bosons through a pseudo-scalar
Higgs boson s--channel exchange.  If, in particular, the mass of the
pseudo-scalar is about twice the DM mass, one expects a large resonant
interaction and potentially a large di-Higgs boson production.  

In both cases however, one also expects a large DM pair annihilation
rate into two $\gamma \, \gamma$, $Z \, Z$, $Z \, \gamma$, $H \,
\gamma$, $H \, Z$ leading to extra $\gamma$--ray lines.  In many
scenarios, these process are related, thus giving interesting
constraints on the model.  However, large branching ratios into
$\gamma \, \gamma$, $Z \, Z$, $Z \, \gamma$, $H \, \gamma$, $H \, Z$
could be detrimental to the searches for a 63~GeV line.  For example, in
`conventional' BSM scenarios such as SUSY, the di-photon final state
is  supposed to be slightly larger than the di-Higgs production
(notably because it is not phase-space suppressed).  Since the
di-photon final state relies on charged loop diagrams, one therefore
expects a large production of charged particles from the DM pair
annihilations at tree-level which poses a problem for the
detectability of the 63~GeV line.  Indeed, if the contribution from
annihilations into $b$-quarks is significant, it is likely that the
line at 63~GeV would be totally swamped by the continuum $\gamma$--ray
emission resulting from the $b$ hadronization, fragmentation and
subsequent decay, with an endpoint energy equal to the DM mass,
$m_\chi \simeq 126$~GeV.

There are several ways out, nevertheless.  For example, if the charged
particles which contribute to the direct photon emission
(loop-suppressed) are all heavier than the DM~\cite{MCcabe}, the DM
pair annihilation into such particles is not kinematically allowed,
thus enabling the  di-Higgs final state to be visible.  In SUSY, this
means that one would have to suppress the t--channel sbottom exchange
diagram and perhaps introduce a singlet-like heavy Higgs boson mostly
coupled to very heavy charged particles~\cite{MCcabe}.  Alternatively,
there could be scenarios where the di-photon and di-Higgs final states
are produced by enhanced box diagrams but in which the sbottom exchanges
are very suppressed so that the production of $b$-quarks at tree-level
is suppressed.  In scenarios with a SM-like Higgs boson and no extra
pseudo-scalar boson, the tree-level production of $b$-quarks is
expected to be velocity-suppressed.  If potential loop/box process,
susceptible to imply $b$-quarks at tree-level, are also suppressed by
the introduction of very heavier mediators, the detectability of the
63~GeV line originating from enhanced box diagrams could be significant.

We also note that in models such as the NMSSM where one can have both a
very heavy ($A$) and very light ($a$) pseudo-scalar Higgs bosons, the
requirement of having a resonant $A$ exchange if $m_\chi \simeq$~126~GeV
(i.e., $m_A= 2 m_\chi \simeq 252$~GeV) implies that one could also
produce at tree-level the $A a$ final state, with $A$ produced at
rest.  The decay of the $A$ into two photons could then generate a
line at 126~GeV which could be confused with the direct (resonant) DM
pair annihilations into two photons.  The dominance of one process
over the other  would mostly depend on the mass difference $|m_A - 2 \,
m_\chi|$ and the strength of the coupling of the neutralino to the
Higgs boson, which itself is constrained by the width of the invisible
Higgs decay channel~\cite{Shrock:1982kd, Pospelov:2011yp,
  Englert:2012wf}.  Such an ambiguity in the origin of a possible line
at $E \simeq 126$~GeV in this framework could be of interest in the
context of the 130~GeV and 111~GeV bumps observed in the {\sl
  Fermi}--LAT data~\cite{Bringmann:2012vr, Weniger:2012tx,
  Rajaraman:2012db, Su:2012ft, Su:2012zg, Bringmann:2012ez,
  Hektor:2012ev, Finkbeiner:2012ez}. 

For candidates with this mass ($m_\chi \simeq$~126~GeV), the condition
of predicting a 63~GeV line from a SM-like Higgs boson produced at 
rest guaranties a final state with two SM-like Higgs bosons.  However 
should such a line be seen, one would have to disentangle it from the
direct annihilations of DM particles with a mass of $m_\chi
\simeq 63$~GeV into two photons.  Also it may be challenging to
disentangle the di-Higgs boson final state from the $H \gamma$ final
state.  These aspects will be discussed in the next section. 

Note that all the final states mentioned above have already been
considered in detail in the literature for generic DM masses (see,
e.g., Refs.~\cite{Bergstrom:1997fh, Bern:1997ng, Jackson:2009kg}).
However, to our knowledge, the $\gamma$--ray signature expected from a
Higgs boson decay produced by a $\sim 126$~GeV DM candidate has not
been studied explicitly\footnote{$\gamma$--ray fluxes have been
  calculated and can be obtained from Ref.~\cite{Cirelli:2010xx} for
  different DM masses for DM annihilations into a pair of SM-like
  Higgs bosons, though.}.  Many authors have exploited the presence of
a single photon in DM pair annihilation final states as a $\gamma$--ray
signature~\cite{Ullio:1997ke, Bergstrom:1997fj, Boudjema:2005hb,
  Perelstein:2006bq, Gustafsson:2007pc, Dudas:2009uq, Mambrini:2009ad,
  Jackson:2009kg, Bertone:2010fn}.  However, the possibility of these
prompt photon lines being accompanied by additional lines due to Higgs
production at rest has not been pointed out.  To our knowledge, the
fact that the DM pair annihilation into two photons could be simply
confused with a Higgs boson (not necessarily SM-like) production, when
$m_{H} \simeq 2 m_\chi$, has not been mentioned in the literature yet.

\subsection{Production mechanisms for $m_\chi \simeq 63$~GeV}

Due to their mass, candidates with $m_\chi \simeq 63$~GeV can only
produce one SM-like Higgs boson at rest in the final state.  The DM
spin is then fixed by the nature of the second particle in the final
state.  The exact final state can also enable one to determine the Higgs
boson production mechanism.  For example, the $H \gamma$ final state
implies that the Higgs boson production must be a loop-suppressed
process since the DM is assumed to be neutral and cannot produce a
photon in the final state without coupling to charged particles
(unless one considers `dipole' DM~\cite{Sigurdson:2004zp}).

Usually one exploits the presence of a single photon in the final
state to look for such a process (see, e.g.,
Ref.~\cite{Jackson:2009kg}).  However,  the corresponding direct
$\gamma$--ray line would appear at very low energy, namely
$E_{\gamma}= m_\chi \, (1 - m_H^2/(4 \, m_\chi^2)) \ll 1$~GeV, to
which {\sl Fermi}--LAT might still be sensitive.  Hence the only line
that is experimentally accessible comes from the Higgs decay at 63~GeV.
Nevertheless, observing such a line may not unambiguously point towards
the production of a Higgs boson: DM pair annihilations into $\gamma \,
\gamma \ $ could also produce a monochromatic line at the same energy
as the Higgs boson decay if the DM mass is about 63~GeV.  Hence, there
could be some confusion about the origin of the line, even though such
a detection would definitely point towards new physics.

In some models, this possible confusion could be solved  by simply
comparing the expected cross sections in different channels.  For
example, in scenarios with photon mixing~\cite{Davoudiasl:2012ag}, the
$Z_d$ s--channel exchange into $\gamma \, H$ would be larger than the
$\gamma \, \gamma$ final state, so a signal at 63~GeV could be
expected.  However there could be tricky situations.  For example, if
$m_\chi \simeq 63$~GeV, both the $\chi \, \chi \, \rightarrow \,
\gamma \, \gamma$  and $\chi \, \chi \, \rightarrow \, H \, \gamma$
process are expected to be very large if they are realized through a
Higgs portal, i.e., $\chi \, \chi \rightarrow H \rightarrow \gamma \,
\gamma, H \gamma$.  The kinematic condition to see a line at $m_\chi
\simeq  m_H/2 \simeq 63$~GeV indeed immediately implies that the $H$
exchange is resonant.  Hence, both final states should be copiously
produced.  If $H$ is the SM Higgs boson, the magnitude of $\chi \,
\chi \, \rightarrow \, \gamma \, \gamma$  versus $\chi \, \chi
\rightarrow H \, \gamma$ is fixed by the ratio of the $t-t-\gamma$
versus the $t-t-H$ couplings and the phase space factor.  Thus, for a
SM-like Higgs boson produced very close to rest, the phase space factor
eventually suppresses a bit the $H \, \gamma$ final state.  Yet,
ultimately one should detect the sum of the two contributions.  

Note that the importance of the $\chi \, \chi \, \rightarrow \, \gamma
\, \gamma$ and $\chi \, \chi \, \rightarrow \, H \, \gamma$ processes
through the SM-like Higgs portal ultimately depends on the mass
difference $\Delta  = 2 m_\chi - m_H$, as well as the $\chi-\chi-H$
coupling.  The latter can be tuned (in fact reduced) to compensate for
the smallness of $\Delta$,  in order to avoid too large a resonant
annihilation effect, although it cannot be arbitrarily large.  The
maximum value of the $\chi \, \chi \rightarrow H \rightarrow \gamma \,
\gamma \ $ cross section is actually set indirectly by the ATLAS and
CMS experiments.  The associated cross section is maximal when both
$\Delta$ becomes smaller than the Higgs boson decay width ($\Gamma_H$)
and the $\chi-\chi-H$ coupling is maximal.  Both are being measured at
LHC through the Higgs visible and invisible decay
width~\cite{Dobrescu:2012td}.  A too large $\chi-\chi-H$ coupling
would make the Higgs decay invisible and be in conflict with SM
predictions.

The above discussion assumes that the DM pair annihilation through  the
Higgs portal cross section is not velocity-dependent.  However, if they
turn out to be suppressed and box diagrams are more important, models
with kinetic mixing might again lead to a larger value of the cross
section for the $H \gamma$ final state (with respect to the $\gamma \, 
\gamma$ final state).

Would such a line be seen, it would remain to be determined whether it
originates from a SM-like Higgs boson decay into two photons or a model
of the type discussed above.  However,  when $m_\chi \simeq 63$~GeV,
the DM pair annihilations into any other channel would produce a
$\gamma$--ray spectrum with energies $E_{\gamma} < m_\chi$.  Hence the
line at $\sim$~63~GeV would not be buried under the continuum spectrum
unlike what could occur for $m_\chi > 63$~GeV, as discussed in the
previous subsection.

\subsection{Production mechanisms for $m_\chi \simeq 109$~GeV}

When the DM mass is about 109~GeV, the $\chi \, \chi  \rightarrow H
Z$ process can occur (for both bosonic and fermionic DM) via a
t--channel DM exchange diagram (if DM can couple directly to the Higgs)
or a s--channel Z exchange diagram. This process can also occur through
box diagrams.

For such a value of the DM mass, both the $Z$ and SM-like $H$ bosons
are produced close to rest and should lead to distinctive signatures.
In addition to the 63~GeV line from the SM-like Higgs boson decay, there
could be a line at $\sim$~109~GeV coming from the DM annihilations
into two photons.  Associated with this case, there could also be a line
at $\sim$~72~GeV from the direct photon in the $H \, \gamma$ final
state if this channel is not suppressed.  The dominance of one over the
other one depends again on the couplings and exact process, while their
visibility essentially depends on the background at these energies.
Note that $\gamma$--ray line at $\sim$~109~GeV from direct
annihilation into two photons could be consistent with the possible
line detected at 111~GeV~\cite{Rajaraman:2012db, Su:2012ft} and could
be used to constrain the DM interactions.

\subsection{Additional remarks \label{sec:subtilities}}

The results displayed in the next section hold independently of
whether the new particle discovered at CERN is the Higgs boson or not.
Since the observed branching ratios are compatible with the SM Higgs
predictions (within 2~$\sigma$), our conclusion regarding whether one
can see a monochromatic line at $\sim 63$~GeV should remain identical.

Some of the Higgs production mechanisms that we discuss in this paper
may be associated with a large spin-independent elastic scattering
cross section with a nucleon and could be ruled out by DM direct
detection experiments.  In particular if the DM has a mass in the
GeV-TeV range, its interactions could be severely constrained by 
the XENON100 experiment~\cite{Aprile:2012nq, Davis:2012hn}.  Since this
requires to specify a model and we intend to set model-independent
constraints, we assume that the underlying DM particle model is
compatible with the results from the latest direct detection
experiments.  However, for concrete models such a compatibility has to 
be checked.

\subsection{Detectability of the line emission and continuum}

\begin{figure}[t]
\centering
\includegraphics[width=1.\linewidth]{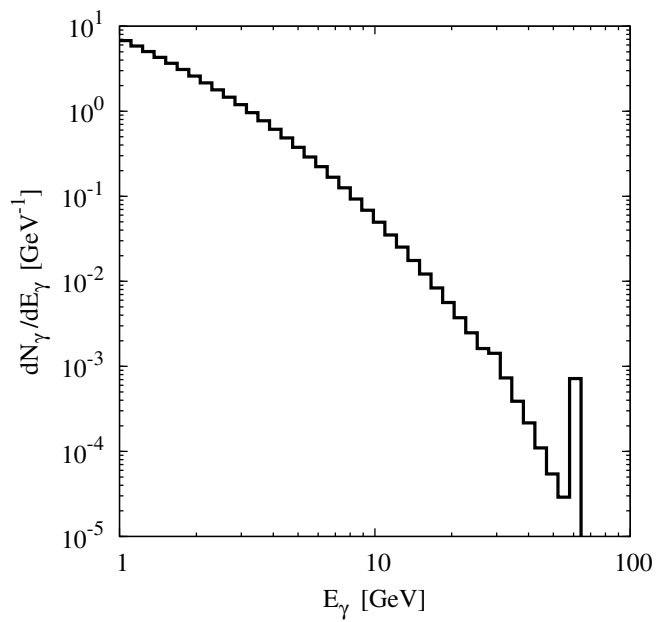}
\caption{Number of photons per GeV produced by the decay of a SM-like
  Higgs boson with mass of 126~GeV produced at rest.  The line at
  63~GeV from $H \rightarrow \gamma \, \gamma$ is suppressed, but
  nevertheless distinguishable from the continuum that arises from the
  Higgs boson decay into the rest of SM particles. Note that the very
 small excess at 30~GeV is due to the prompt photon coming from $H
 \rightarrow Z \, \gamma$.}
\label{fig:hdecay_rest}
\end{figure}

The $\gamma$--ray emission subsequent to Higgs production typically
occurs from the Higgs boson decay into, e.g., $\gamma \, \gamma$, $b
\, \bar{b}$, etc.  Since all the channels have very well-known branching
ratios, the $\gamma$--ray flux can be predicted quite accurately
(albeit astrophysical uncertainties).

Predictions depend on the photon energy spectrum $dN_{\gamma}/dE_\gamma$
associated with the Higgs boson decay.  Typically, for a Higgs boson
of about 126~GeV produced at rest, one expects a smooth spectrum (due
to dominant decay into $b \, \bar{b}$) plus a monochromatic line due to
$H \rightarrow \gamma \, \gamma$~\cite{Ellis:1975ap, Djouadi:1993ji,
  Dittmaier:2012vm}.  In the SM, (for $m_H = 126$~GeV) the Higgs boson
decay into $\gamma \, \gamma$ is suppressed by a factor of $\sim 4
\times 10^{-3}$ with respect to the $b \, \bar{b}$ final
state~\cite{Dittmaier:2012vm}, so one may think that the $\gamma$--ray
line is hidden by the continuum.  However, channels such as $b \,
\bar{b}$ emit photons at lower energies than $E = m_H/2$ (owing to
final state radiation, hadronization, fragmentation and decay).  As a 
result, even though the flux associated with the monochromatic line is
meant to be suppressed, in principle it could be distinguishable from
the continuum emission.  In order to compute the $dN_{\gamma}/dE_\gamma$
spectrum, we use PYTHIA 6.4~\cite{Sjostrand:2006za}, where we set the
branching ratio for $H \rightarrow \gamma \, \gamma$ to $2.28 \times
10^{-3}$~\cite{Dittmaier:2012vm}~\footnotemark.  The result is
displayed in Fig.~\ref{fig:hdecay_rest}.  Clearly, the monochromatic
line appears to be distinguishable from the smooth spectrum, even
though it is suppressed.

Now, we estimate the associated flux from DM annihilations (an
analogous analysis could be performed for decaying DM) around the
galactic center and compare it to the current {\sl Fermi}--LAT data.
We will assume a generic DM candidate, with a thermal average of the
annihilation cross section times the relative velocity of $\langle
\sigma \, v \rangle \equiv \langle \sigma \, v_{DM DM \rightarrow H +
  (\gamma, \, Z, \, H)} \rangle = 3 \times 10^{-26}
\textrm{cm}^3/\textrm{s}$,  where in each case we consider that the 
only annihilation channel is $H \, H$, $H \, \gamma$ or $H \, Z$.  

The differential flux of prompt $\gamma$--rays generated from DM
annihilations in the smooth DM halo from a direction within a solid
angle $\Delta\Omega$ is given by~\cite{Bergstrom:1997fj}
\begin{equation}
\frac{d\Phi_{\gamma}}{dE_\gamma} = \eta \, \frac{\langle\sigma \, 
  v\rangle}{m_\chi^2} \, \frac{dN_{\gamma}}{dE_{\gamma}} \, 
\frac{1}{8 \, \pi} \, \int_{\Delta\Omega}d\Omega \,
\int_\text{los}\rho\big(r(s,\Omega)\big)^2 \, ds\,,  
\label{Eq:promptflux}
\end{equation}
where $dN_{\gamma}/dE_{\gamma}$ is the differential $\gamma$--ray yield,
$\eta$ is a symmetry factor which for Majorana DM is equal to 1 and
1/2 if DM is not a self-conjugate particle, $\rho(r)$ is the DM
density profile and $r$ is the distance from the galactic center.  The
spatial integration of the square of the DM density profile is
performed along the line of sight within the solid angle of
observation $\Delta\Omega$.  More precisely, $r = \sqrt{R^2_\odot -  2
  s R_\odot \cos \psi + s^2}$, and the upper limit of integration is
$s_{\rm max} = \sqrt{(R_{\rm MW}^2 - \sin^2 \psi R^2_\odot)} + R_\odot
\cos \psi$, where $\psi$ is the angle between the direction of the
galactic center and that of observation and $R_\odot$ is the distance
from the Sun to the galactic center.  Being the contributions at large
scales negligible, the choice of the size of the Milky Way halo,
$R_{\rm MW}$ is not crucial.

Thus, the flux of DM annihilations can be written as
\begin{widetext}
\begin{equation}
\frac{d\Phi_{\gamma}}{dE_\gamma} =  9.27 \cdot 10^{-9} \,
\mathrm{cm^{-2}\,s^{-1}} \times  \eta \times
\frac{dN_{\gamma}}{dE_{\gamma}} \times \left(\frac{\int J(\psi)
  d\Omega}{20.5 \, \text{sr}}\right) \, \left( \frac{m_\chi}{100 \,
  \rm{GeV}} \right)^{-2} \, \left(\frac{\langle \sigma \, 
  v\rangle}{3\cdot 10^{-26} {\mathrm{cm^3/s}}}\right) \,
\left(\frac{\rho_\odot}{0.386 \, \rm{GeV/cm^3}}\right)^2 \,
\left(\frac{R_\odot}{8.25 \, \rm{kpc}}\right)  ~, 
\label{Eq:totpromptflux}
\end{equation}
\end{widetext}
with the dimensionless quantity $J(\psi)$ defined as
\begin{equation}
J(\psi) = \frac{1}{R_\odot\,\rho_\odot^2} \, \int_\text{los}
\rho\big(r(s,\Omega)\big)^2\ ds \, ~,  
\label{Eq:J} 
\end{equation}
where for the distance from the Sun to the galactic center and for the
local DM density we use $R_\odot = 8.25$~kpc and $\rho_\odot =
0.386$~GeV/cm$^3$, respectively~\cite{Catena:2009mf}.

\begin{figure}[t]
\centering
\includegraphics[width=1.\linewidth]{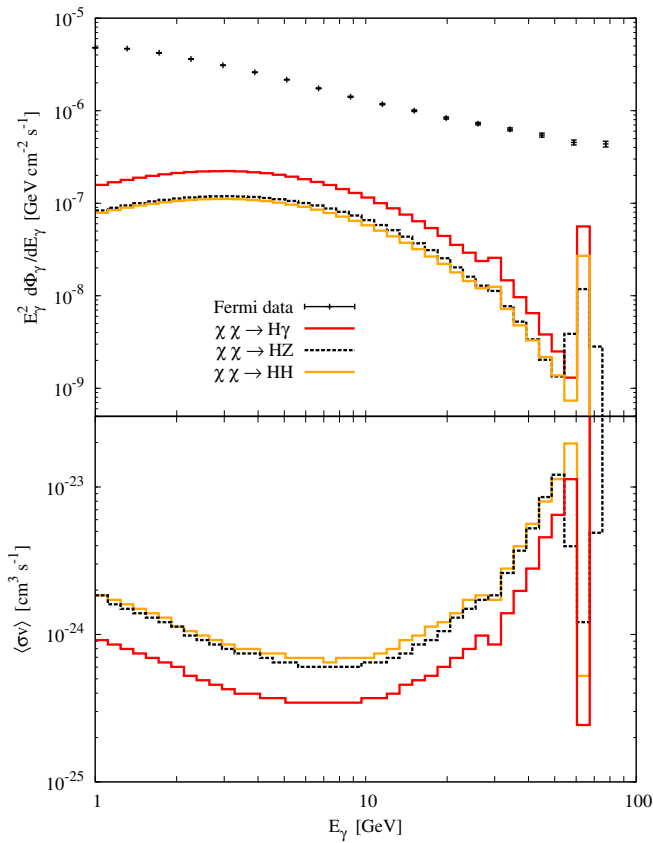}
\caption{{\sl Upper panel:} Potential $\gamma$--ray flux from the
  galactic center due to DM annihilating into $H \, \gamma$ (upper
  red line), $H \, Z$ (black dotted line) and $H \, H$ (orange line),
  when the Higgs is produced very close to rest, i.e., for $m_\chi =
  63$~GeV, $109$~GeV and~$126$~GeV, respectively.  The results are for
  an Einasto profile for a $20^o\times20^o$ squared region around the
  galactic centre and for $\langle \sigma \, v \rangle = 3 \times
  10^{-26} \rm{cm^3/s}$.  The dots represent the {\sl Fermi}--LAT data
  points in that region for about 4 years.  {\sl Lower panel:} Values
  of the annihilation cross section for each case for which the signal
  flux would be equal to the background flux. Note that the
  astrophysical sources are not subtracted from the data points.
}
\label{fig:spectrum_hdecay_rest}
\end{figure}

Although for some DM density profiles, the integration of $J(\psi)$ in
the solid angle of observation can be done
analytically~\cite{Lavalle:2009fu}, here we consider an Einasto 
profile~\cite{Einasto}, for which there is no analytical solution, and
compute it numerically.  This density profile is parametrized as
\begin{equation}
 \rho(r) = 0.193 \, \rho_\odot \, \exp\left[ -\frac{2}{\alpha} \left(
   \left(\frac{r}{r_s} \right)^\alpha - 1 \right) \right]\,,\qquad
 \alpha = 0.17 ~, 
\label{Eq:einasto}
\end{equation}
where $r_s = 20$~kpc is a characteristic length.~\footnotetext{Note
  that the default value in PYTHIA 6.4 is $3.45 \times 10^{-3}$.  Also
  notice that the best fit for the LHC measurement is 2.9 times the SM
  value and, at 95\% confidence level, it could be up to 5.4 times the
  SM one~\cite{Corbett:2012dm} (see also Refs.~\cite{Carmi:2012yp,
    Azatov:2012bz, Espinosa:2012ir, Montull:2012ik}).}  

Following Refs.~\cite{Serpico:2008ga, Jeltema:2008hf, Pieri:2009je,
  Bernal:2010ip, Bernal:2011pz}, we consider a $20^o\times20^o$
squared region centred on the galactic center, for which $\int J(\psi)
d\Omega = 20.5 \, \textrm{sr}$.  In
Fig.~\ref{fig:spectrum_hdecay_rest} we compare the expected flux from
this region and compare it with the {\sl Fermi}--LAT data.  To obtain
the measured flux, we take the {\it Fermi}--LAT data obtained from
August 4, 2008 to October 1, 2012.  We extract the data from the {\it
  Fermi} Science Support Center archive~\cite{Fermidata} and select
only events classified as \texttt{CLEAN}.  We use a zenith angle cut 
of 105$^\circ$ to avoid contamination by the Earth's albedo and the
instrument response function \texttt{P7CLEAN\_V6}.

In the upper panel of Fig.~\ref{fig:spectrum_hdecay_rest} we show the
$\gamma$--ray spectra for three different annihilation channels, $H \,
\gamma$ (upper red line), $H \, Z$ (black dotted line) and $H \, H$
(orange line), in which the Higgs is produced very close to rest.  The
DM mass for each case is $m_\chi = 63$~GeV, $109$~GeV and~$126$~GeV,
respectively.  As can be seen from the plot, the fluxes for the three
cases are very similar, but the $H \, \gamma$ final state is slightly
more visible than the two others\footnote{Let us stress again that the
  $\gamma$--rays in this case are only those coming from Higgs decay.
  For $m_\chi \simeq 63$~GeV, the direct photon lies at energies well
  below detection threshold for {\sl Fermi}--LAT.}, mainly because of
the lower value of the DM mass in this case.  Since the flux scales
linearly with the cross section, these lines emerge from the
$\gamma$--ray background when the associated production cross section
is greater than $\langle \sigma \, v \rangle \sim 2.5 \, (5) \times
10^{-25} \rm{cm^3/s}$ for $H \, \gamma$ ($H \, H$), thereby ruling out
a Higgs boson production cross section larger than this value.  This
can be seen from the lower panel of
Fig.~\ref{fig:spectrum_hdecay_rest}, where we show the value of
$\langle \sigma \, v \rangle$ for which the signal would be equal to
the observed background.  Interestingly enough, for the case of DM
annihilations into $H \, \gamma$ or $H \, H$, producing Higgs at rest,
the $\gamma$--ray line from the very suppressed $H \rightarrow \gamma \,
\gamma$ channel (see Fig.~\ref{fig:hdecay_rest}), is expected to
provide a more restrictive limit than the dominant continuum.  

The limits that we sketch are very conservative as they assume no
background from astrophysical sources.  A dedicated search for Higgs
boson decay lines would require to account for the background modeling
and to optimize the detection window~\cite{Bringmann:2012vr,
  Weniger:2012tx, Rajaraman:2012db, Su:2012ft, Su:2012zg,
  Bringmann:2012ez, Hektor:2012ev, Finkbeiner:2012ez}. However here we
simply want to illustrate the potential detectability of these
lines.  Note that our limits are in agreement with the detailed {\sl
  Fermi}--LAT searches of $\gamma$--ray
lines~\cite{Ackermann:2012qk}.  These  were obtained by the {\sl
  Fermi}--LAT analysis for $m_\chi \simeq 63$~GeV and $\chi \, \chi
\rightarrow \gamma \, \gamma$ can be directly compared to the ones
presented here for $\chi \, \chi \rightarrow H \, \gamma$ and $m_\chi
\simeq 63$~GeV.  While the {\sl  Fermi}--LAT limit is $\langle \sigma
\, v \rangle \sim 3 \times \ 10^{-28} \rm{cm^3/s}$ ({\sl cf.} Fig.~15 in
Ref.~\cite{Ackermann:2012qk}), we obtain $\langle \sigma \, v \rangle
\sim 2.5 \times 10^{-25} \rm{cm^3/s}$, the $\sim 10^{-3}$ difference
coming from the branching ratio for $H \rightarrow \gamma \, \gamma$.
Similarly, for the case of $\chi \, \chi \rightarrow H \, H$ and
$m_\chi \simeq 126$~GeV, the limit obtained from the $\gamma$-ray line
from Higgs decay is just a factor of 2 weaker than that for $\chi \,
\chi \rightarrow H \, \gamma$ and $m_\chi \simeq 63$~GeV (explained as
a factor of 2 in favour of $H \, H$ due to having two Higgs bosons and a
factor of 4 in favour of $H \, \gamma$ due to the factor of two in the
DM mass).

In DM models where there is a correlation between the di-photon and $H
\, \gamma$, $H \, Z$ and/or $H \, H$ final states, the ratio of the
flux associated with the prompt $\gamma$-ray line to that of the Higgs
boson decay line can be used to test the model.  In particular when
$m_{\chi} \simeq 126$~GeV, one expects the following ratio
$\frac{\phi_{\gamma \, \gamma}}{\phi_{H \, H}} =~\frac{1}{BR_{H
    \rightarrow \gamma \, \gamma}} \times \frac{\sigma \, v_{\gamma \,
    \gamma}}{\sigma \, v_{H \, H}}$.
   
In the absence of evidence for a specific DM model and a precise
correlation between these two final states, searching for the Higgs
decay line could allow us to obtain a constraint on the DM-Higgs boson
interactions.  The main difficulty associated with these searches
consists in removing the astrophysical background sources but 
these searches are worthwhile, as they could reveal new physics and
point towards models with multiple scalar and pseudo-scalar Higgs
bosons with large DM-Higgs couplings, for example.

\section{Boosted Higgs and multiple Higgs bosons
  scenarios \label{sec:boost}}

We can now investigate the case of boosted Higgs production and
multiple Higgs scenarios.

\subsection{Boosted Higgs boson}

The Higgs boson decay line considered in the previous section is now
replaced by a broad excess which shows up as a less prominent feature.
For $\chi \, \chi \rightarrow H \, H$, this box-shaped part of the
spectrum is a particular case of those studied in
Ref.~\cite{Ibarra:2012dw}.  However, in the cases discussed here, this
broad excess is accompanied by a smooth spectrum from the Higgs decay
into all other possible channels plus a possible line due to prompt
photon emission in the $H \, \gamma$ final state.

\begin{figure}[t]
\centering
\includegraphics[width=1.\linewidth]{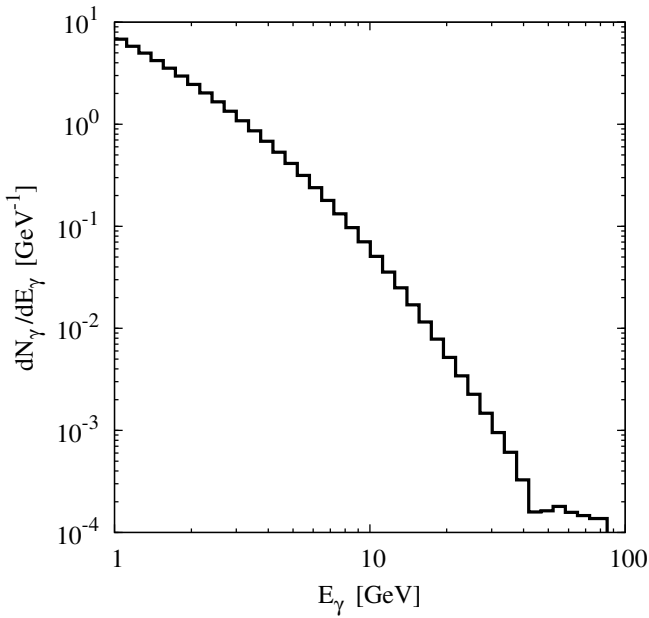}
\caption{Number of photons per GeV expected from the decay of a
  boosted SM-like Higgs boson with $m_H = 126$~GeV produced with an
  energy $E_H = 130$~GeV.  As one expects, due to the boost, there
  is no line at 63~GeV from $H \rightarrow \gamma \gamma$, but one can
  nevertheless see a broad (box-shaped) emission.} 
\label{fig:hdecay_boosted}
\end{figure}

\begin{figure}[t]
\centering
\includegraphics[width=1.\linewidth]{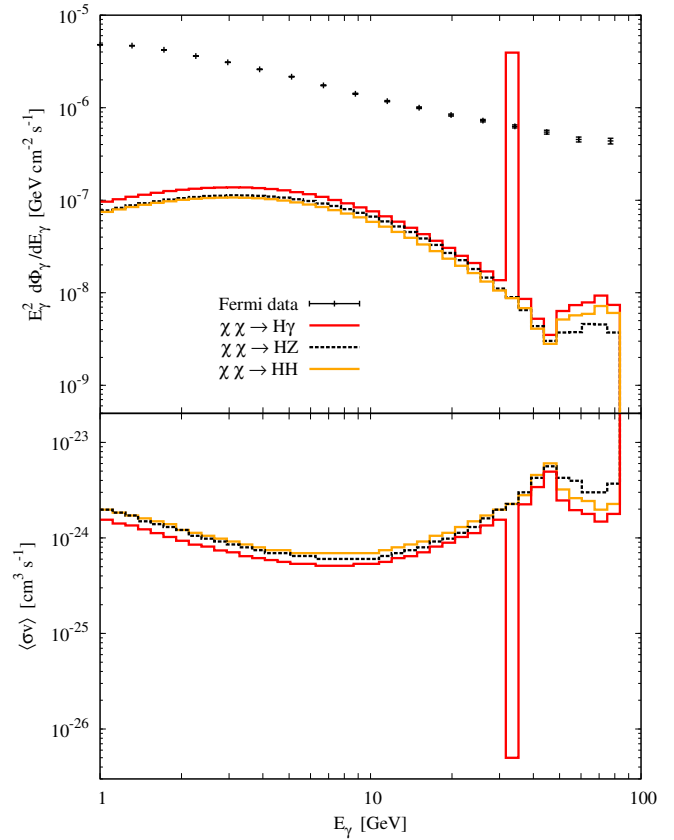}
\caption{Same as Fig.~\ref{fig:spectrum_hdecay_rest}, but for DM
  masses such that the Higgs boson is produced with an energy $E_H
  \simeq 130$~GeV, i.e., for $m_\chi = 81$~GeV ($H \, \gamma$, upper
  red line), 111~GeV ($H \, Z$, black dotted line) and 130~GeV ($H \,
  H$, orange line).}
\label{fig:spectrum_hdecay_boosted}
\end{figure}

These features are illustrated in Fig.~\ref{fig:hdecay_boosted}, where
the $\gamma$--ray spectrum due to Higgs decay for a Higgs boson ($m_H
= 126$~GeV) produced with an energy $E_H \simeq 130$~GeV is depicted.
Over the continuum from the other Higgs decay channels, a bump at
$\sim$~60~GeV, corresponding to the Higgs boson decay into two
photons, can still be distinguished.  Below 10~GeV, the continuum is
two orders of magnitude (or more) brighter than the line, so the limit
on the Higgs boson production, for DM masses for which the Higgs boson
is boosted, is actually obtained from the continuum rather than from
the broad excess at $E_\gamma \sim$~60~GeV.  This can be seen in
Fig.~\ref{fig:spectrum_hdecay_boosted}, which is analogous to
Fig.~\ref{fig:spectrum_hdecay_rest}, but now for $m_\chi = 81$~GeV ($H
\, \gamma$), 111~GeV ($H \, Z$) and 130~GeV ($H \, H$), such that, for
all these cases, the produced Higgs has an energy close to 130~GeV.

For the $H \, \gamma$ final state, note that there is a $\gamma$--ray
line emitted at 32~GeV, in addition to the box-shaped spectrum at
$E_\gamma \sim$~50--80~GeV and the continuum from Higgs decays.  This
line originates from the prompt $\gamma$ in the final state and
provides the most stringent bound on Higgs boson production cross
section.  Actually, in the case of $\chi \, \chi \rightarrow H \,
\gamma$, the prompt $\gamma$--ray is always in the energy window
accessible by {\sl Fermi}--LAT if the Higgs is not produced very close 
to rest.  Using the {\sl Fermi}--LAT  data for this annihilation channel
and for $m_\chi \simeq 81$~GeV, we obtain a limit of about 
$\langle \sigma \, v \rangle \lesssim 4 \times 10^{-27}
\rm{cm^3/s}$.  This is comparable to the $\gamma$--ray line limits
obtained by {\sl Fermi}--LAT for $\chi \, \chi \rightarrow \gamma \,
\gamma$ with $m_\chi \simeq 32$~GeV, that is $\langle \sigma \, v
\rangle \lesssim 2 \times 10^{-28} \rm{cm^3/s}$ ({\sl cf.} Fig.~15 in 
Ref.~\cite{Ackermann:2012qk}), after correcting the $\chi \, \chi
\rightarrow H \, \gamma$ cross section limit by a factor of $(1/2) \,
(32/81)^2$ to account for the fact that there is only one prompt
photon in the $H\gamma$ final state with respect to $\gamma\gamma$ and
that the DM mass is different.

\subsection{Multiple Higgs bosons scenarios}

In minimal SUSY models, in addition to a SM-like Higgs, one expects a
heavier CP-even Higgs ($H_2$) and a heavier CP-odd Higgs (A).  If the
heavier CP-odd Higgs boson mass is about $2 m_\chi$, annihilations
into $\gamma \, \gamma$ through CP-odd Higgs portal could be resonant 
and produce a line at $m_{\chi}$.  In fact, this process has been
proposed to explain the bump at 130~GeV in the {\sl Fermi}--LAT
data~\cite{LopezHonorez:2012kv, Das:2012ys, Wang:2012ts}.  In these
configurations, the $A \, \gamma$ and $H_2 \, \gamma$ final states
might be possible too, leading to the production of a CP-odd Higgs
boson on-shell or slightly boosted CP-even $H_2$ if $m_{H_2} \simeq
m_A$.  These final states should be slightly suppressed with respect
to the $\gamma \, \gamma$ final states due to the phase-space
suppression factor, but would still contribute to the $\gamma$--ray
data at $E_\gamma = m_\chi$.

In the NMSSM, final states such as $A \, a$ and $H_2 \, a$  may be
possible too, with $a$ a second pseudo-scalar Higgs boson which can be  
light and $A, H_2$ two heavy Higgs bosons~\cite{Gunion:2012gc}.  Such
final states could lead to the production of Higgs bosons produced at
rest when $2 m_{\chi} \simeq m_{A,H_2} + m_a$ and could be resonant
when $m_a \ll m_A$.  The same process could be in fact relevant for low
DM mass scenarios such as those discussed in
Ref.~\cite{Vasquez:2012hn}.

\section{Conclusions \label{sec:conclusions}}

In this Letter, we have considered the $\gamma$--ray signatures from the
decay of a Higgs boson produced in our galactic halo from DM
annihilations.  We have considered, in particular, the case where the
Higgs boson is SM-like (with a mass of 126~GeV and SM branching
ratios) and showed that the Higgs boson production cross section for
annihilating DM particles with masses $m_\chi \simeq$~63~GeV, 109~GeV
and 126~GeV (Higgs produced very close to rest), cannot exceed
$\langle \sigma \, v \rangle \sim \textrm{few} \times 10^{-25}
\rm{cm^3/s}$.  The limit is in fact mostly driven by the $\gamma$--ray
line from $H \rightarrow \gamma \, \gamma$.  These results can be
trivially generalised to other Higgs boson masses (as relevant in BSM 
models with multiple Higgs bosons and Higgs mass spectrum such as the
NMSSM) leading to different DM scenarios.

We have also considered the case of a slightly boosted Higgs boson and
shown that the associated signature would exhibit a broad
(box-shaped) $\gamma$--ray excess.  However, the continuum associated
with the other Higgs boson decay modes and to the second particle in the
final state would lead to a brighter $\gamma$--ray emission, which can
be used to constrain the Higgs boson production cross section.
Focusing in particular on the $H \, \gamma$ final state for a SM-like
Higgs boson produced with an energy $E_H=130$~GeV, we find that the
Higgs boson production cross section cannot exceed $\sim 4 \times
10^{-27} \rm{cm^3/s}$.

Therefore, we have obtained a simple estimate for the limit on the
Higgs boson production cross section that is independent of any other DM
annihilation channels and demonstrates that performing Higgs boson
decay line searches could be useful to probe the Higgs boson dark
couplings (i.e., couplings to DM particles).  This must be compared to
the limits set on the invisible Higgs boson decay branching ratios
obtained by using LHC measurements ({\sl cf.}, for example,
Ref.~\cite{Englert:2012wf}), but the two approaches (collider and
indirect detection searches) are complementary.

\section*{Acknowledgments}
We would like to thank M.~Cirelli, M.~Dolan, G.~G\'omez-Vargas and
C.~McCabe for useful discussions.  NB is supported by the DFG
TRR33 `The Dark Universe'.  CB and SPR thank the Galileo Galilei
Institute for Theoretical Physics for its hospitality.  JS and CB are
supported by the ERC advanced grant `DARK' at IAP, Paris.  SPR is
partially supported by the Portuguese FCT through CERN/FP/123580/2011
and CFTP-FCT UNIT 777, which are partially funded through POCTI (FEDER)
and by the Spanish Grant FPA2011-23596 of the MINECO.  TT is supported by
the European ITN project (FP7-PEOPLE-2011-ITN,
PITN-GA-2011-289442-INVISIBLES).  Numerical
computation in this work was partially carried out at the Yukawa
Institute Computer Facility.

\bibliography{higgs}

\end{document}